\documentclass[12pt]{iopart}
\begin{document}

\title[]{Detection of cosmic neutrino clustering by cosmic ray
spectra}
%{Preparing an article for publication in an Institute of Physics
%Publishing journal using \LaTeXe}

\author{W-Y. P. Hwang\dag\ and Bo-Qiang Ma\ddag
\footnote[3]{To whom correspondence should be addressed
(mabq@phy.pku.edu.cn). Mailing address: Department of Physics,
Peking University, Beijing 100871, China.} }

\address{\dag\ Department of Physics, National Taiwan University,
Taipei, Taiwan 10617}

\address{\ddag\ Department of Physics, Peking University, Beijing
100871}

\begin{abstract}
We propose a method to investigate the scenario that cosmic relic
neutrinos are highly clustered around stars and galaxies, or
dark-matter clusters, rather than uniformly distributed in the
universe. Such a scenario can be detected or constrained by the
interaction of high energy cosmic ray protons and nuclei with the
cosmic neutrinos. There should be observable signature in the
energy spectra of cosmic ray protons and nuclei for a neutrino
clustering factor beyond $10^{13}$. We provide a relation on the
signature onset positions between proton and nuclei spectra, and
discuss possible support from existing experiments. It is also
suggested that the relative abundance of cosmic ray nuclei may
detect or constrain the cosmic neutrinos with smaller clustering.

\end{abstract}

%Uncomment for PACS numbers title message
%\pacs{00.00, 20.00, 42.10}
% PACS, the Physics and Astronomy
                             % Classification Scheme.

% Uncomment for Submitted to journal title message

%\submitto{\NJP}

\vfill

\centerline{Published in New Journal of Physics 7~(2005)~41.}

\centerline{The open-access journal for physics:
http://www.njp.org}

 \vfill
%{\NJP{2005}{7}{41}}
% Comment out if separate title page not required

\maketitle

From the standard models of cosmology and particle physics, the
history of our universe can be traced back to about $10^{-11}$
seconds after the Big Bang~\cite{BB} using well-established
physical laws. At that time the universe was a mixture of photons,
leptons, and quarks at thermal equilibrium with temperature of $T
\approx 300$~GeV. Neutrinos (antineutrinos) decoupled from the
thermal universe at the universe age of 1 second with $T \approx
1$~MeV, and photons decoupled from the baryonic matter at the age
of $3\times 10^{5}$ years with $T \approx 0.3$~eV. Afterward,
neutrinos, photons, and baryonic matter are supposed to evolve
independently as the universe expands. The mean number densities
of these three components in the present universe are well
estimated as $n_{\nu}=(3/22)n_{\gamma} \approx
56$~$\mathrm{cm}^{-3}$ per species of neutrinos (antineutrinos),
$n_{\gamma}=413$~$\mathrm{cm}^{-3}$ for photons, and $n_{B} \sim 5
\times 10^{-10} n_{\gamma} \approx 2\times
10^{-7}$~$\mathrm{cm}^{-3}$ for baryonic matter. From well
established observations of cosmic microwave background (CMB)
\cite{CMB}, we know that photons are distributed in uniform
density with small fluctuation of the order $ 10^{-5}$
\cite{COBE}. However, baryonic matter are distributed in clusters
as galaxies and stars, as a result of gravitational attraction.
Taking the sun as an example, the baryon number density is
$n_{\odot}=8.4\times 10^{23}$ $\mathrm{cm}^{-3}$, with a
clustering factor of $n_{\odot}/n_{B} \sim 4\times 10^{30}$
compared with the mean baryon number density of the universe.
Thus, photons and baryonic matter have distinct patterns of
distributions in the universe.

It is completely unknown whether neutrinos are distributed in
uniform density in our universe or gathered as clusters around the
stars and galaxies. One would expect a uniform distribution in
analogy to photons, if the neutrino mass is exact zero. However,
it has become evident that neutrinos do have nonzero masses from
the oscillation measurements \cite{neutrino-os}.
% and tritium experiments \cite{triton}.
So one may speculate that neutrinos are clustered around the stars
and galaxies, or dark-matter clusters, in a pattern between
photons and baryonic matter. In analogy to the air atmosphere
surrounding the earth, there could be also neutrino atmosphere or
clouds around stars and galaxies, or dark-matter clusters. As
neutrinos interact with baryonic matter very weakly, the neutrino
clouds can penetrate through stars and galaxies. As far as we
know, there is no convincing measurement of neutrino clouds yet.
It is thus necessary to design practical methods for the detection
or constraint of the relic cosmic neutrino background (C$\nu$B),
perhaps through detection of neutrino clouds of certain form. We
will indicate in this work that systematic studies of high energy
cosmic ray energy spectra of protons and nuclei may serve for such
purpose.

We first look at the energy spectrum of high energy cosmic ray
protons, and check the influences due to interact with CMB photons
and C$\nu$B neutrinos in their propagation. We know that there
should be a big suppression in the spectrum, known as
Greizen-Zatsepin-Kuzmin (GZK) cutoff \cite{GZK}, at threshold
$E^{\pi}_{p}\approx %1.1 \times
10^{20}$~eV due to pion production by incident protons on CMB
photons, $p+\gamma\to \pi+ N$. Below that, there is also a
suppression caused by electron-position pair production through
$p+\gamma\to e^{+}+e^{-}+p$ at threshold
$E^{e^{+}e^{-}}_{p}=7.5\times 10^{17}$~eV. These two processes
will produce two ``knees" due to suppressions in the proton energy
spectrum above the corresponding thresholds, and one would also
expect to see two ``bumps" in the spectrum just below the
thresholds owing to the pileup of protons from higher energies.
Compton scattering $p +\gamma \to \gamma+p$ does not require an
energy threshold, and the energy loss is also small for each
collision, therefore it does not cause any non-smooth feature in
the spectrum. We proceed to consider the interaction of the cosmic
ray protons on the C$\nu$B neutrinos,
\begin{equation}
p+\bar{\nu}_{e} \to e^{+} +n. \label{pnu}
\end{equation}
When the proton with 4-momentum $p=(E,{\mathbf p})$ interacts with
the neutrino with 4-momentum $k=(\epsilon_{\nu}, {\mathbf k})$,
and composes into a system with center of mass energy squared $S$,
we have
\begin{equation}
E=\frac{S-m_p^2-m_{\nu}^2 }{2\epsilon_{\nu}
\left(1-\sqrt{1-m_p^2/E^2}(\left|{\mathbf
k}\right|/\epsilon_{\nu})\cos \theta\right)},
\end{equation}
where $\theta$ is the angle between ${\mathbf p}$ and ${\mathbf
k}$, $m_p$ is the proton mass, and $m_{\nu}$ is the neutrino mass.
The energy of the proton $E$ must be very large, so that we have
\begin{equation}
E \approx \frac{S-m_p^2}{2\epsilon_{\nu} \left(1- (\left|{\mathbf
k}\right|/\epsilon_{\nu}) \cos \theta \right)}.
\end{equation}
The threshold energy is
\begin{equation}
E^{{\nu}_e}_{p} \approx \frac{(m_n+
m_{e})^2-m_p^2}{2\epsilon_{\nu}(1+\left|{\mathbf
k}\right|/\epsilon_{\nu})}. \label{Epnu}
\end{equation}
For zero mass neutrinos we have $\epsilon_{\nu}
=\left(4/11\right)^{1/3}\epsilon_{\gamma}=4.54\times 10^{-4}$~eV
and $\left|{\mathbf k}\right|/\epsilon_{\nu}=1$, thus we get
$E^{{\nu}_e}_{p}=1.87 \times 10^{18}$~eV, which is between the
pion and $e^{+}e^{-}$ production thresholds. Assuming the neutrino
energy $\epsilon_{\nu} \approx m_{\nu_{e}} \approx 1$~eV, then
$\left|{\mathbf k}\right|/\epsilon_{\nu} \approx 0$, thus we get
$E^{{\nu}_e}_{p}=1.69\times 10^{15}$~eV, which is well below the
pion and $e^{+}e^{-}$ productions on CMB photons. Adopting the
upper limit of neutrino mass $m_{{\nu}_e}\le 2.2$~eV from tritium
experiments \cite{triton}, we get $E^{{\nu}_e}_{p} \ge 7.70\times
10^{14}$~eV. Thus the threshold position $E^{{\nu}_e}_{p}$ ranges
from $7.70\times 10^{14}$ to $1.87 \times 10^{18}$~eV. In
principle, the process (\ref{pnu}) should also produce a ``knee"
in the proton cosmic ray spectrum above the threshold and a
``bump" below, as suggested by Wigmans~\cite{Wig00}. But the
magnitudes might be very small, we need to check the constraints
on C$\nu$B for any observable consequences.

The mean energy of relic neutrinos is too small for interaction
with the ordinary baryonic matter on earth, such that a high
condensed neutrino clustering in cosmic background is not in
conflict with our conventional knowledge. If relic neutrinos are
condensed with a clustering factor $c_{\nu}$ compared to the mean
number density, the number density should be
$n_{\nu}=56c_{\nu}$~$\mathrm{cm}^{-3}$. The cross section of
reaction (\ref{pnu}) is of the order
\begin{equation}
\sigma %\approx %\frac{G_F^2 S}{\pi}
%G_F^2 S/\pi \sim 1.5 \times 10^{-38}~{\mathrm{cm}}^2
\sim 10^{-43}~\mbox{cm}^2 \left(\frac{m_{\nu}}{\mbox{eV}}\right)^2
\left(\frac{E_p^{\nu}}{\mbox{PeV}}\right)^2 \sim 10^{-42}
~\mbox{cm}^2,
\end{equation}
for $m_{\nu} \sim 0.5~\mbox{eV}$ and $E_p^{\nu}=4\times
10^{15}~\mbox{eV}$ calculated later on. Therefore the mean free
path for a proton to propagate in C$\nu$B is
\begin{equation}
\lambda_{p}^{{\nu}_e}=%\frac{1}{n_{\bar{\nu}_e} \sigma }
\frac{1}{n_{{\nu}_e} \sigma}  \sim \frac{4 \times 10^{15}}
{c_{{\nu}_e}}~{\mathrm{Mpc}} .
\end{equation}
A clustering factor of $c_{{\nu}_e}\sim 10^{13}$ would therefore
correspond to a mean free path $\lambda_{p}^{{\nu}_e} \sim 400$
~Mpc. As the protons with energies below the GZK cutoff and
$e^+e^-$ production threshold can come from any source as large as
the Hubble length, we thus predict an observable effect in the
energy spectrum of proton cosmic rays, provided that relic
neutrinos are clustered with a factor of the order $10^{13}$ or
larger. The cosmic neutrino clustering may occur around our
galaxy, or around dark-matter clusters or supermassive black holes
which are possible origin for the ultrahigh energy cosmic ray
acceleration \cite{BG99}. The sizes of these neutrino clusterings
are finite, might be a few orders larger than the sizes of the
galaxies and dark-matter clusters. If we take the neutrino
clustering size as the same order of a typical galaxy cluster
size, i.e., 20~Mpc, then we will need a larger clustering factor
of $c_{\nu}=10^{14}$ for the neutrino clustering to be observable.
In this case, 30\% of the primary cosmic ray protons and nuclei
from outside the galaxy cluster would experience at least once
interaction with cosmic neutrinos.

We briefly review the available studies on the relic neutrino
clustering. The clustering of relic neutrinos is suggested by
Weiler in his proposal of using $\nu +\bar{\nu}\to Z$ to detect
the relic neutrinos \cite{Weiler}. In order to explain the
ultrahigh energy cosmic ray events above the GZK cutoff, the
Z-bursts are suggested as the possible source, and the clustering
factor for the cosmic neutrino background is adopted from a range
of $10^{2\to 7}$ \cite{Weiler,FMS,Yos98} to $10^{8 \to 14}$
\cite{McK01} to generate the required flux. A large neutrino
clustering with order of $10^{13}$ is also considered
\cite{Moh97,Rob91} to explain an anomaly in the tritium beta decay
spectrum \cite{Rob91,Bel95} as from a background contribution of
${\nu}_e+ {^3{\mathrm{H}}} \to e^{-}+ {^3{\mathrm{He}}}$. It is
suggested by Wigmans \cite{Wig00} that the knee of all-particle
cosmic ray spectrum at energy $4\times 10^{15}$~eV is due to the
process (\ref{pnu}) with a neutrino mass $m_{{\nu}_e} \approx
0.4$~eV, provided with a dense neutrino clustering. There is also
a theoretical calculation of the neutrino clouds as a result of
very weakly interaction between neutrinos by the exchange of a
very light scalar boson \cite{Ste98}. Therefore a neutrino
clustering factor in a range $10^{2 \to 14}$ is by no means
nonsense; it is still $16 \to 28$ orders smaller than the baryonic
matter clustering factor $4\times 10^{30}$. It is important to
sharpen the range, and to reveal or to rule out the neutrino
clustering with convincing evidence. We will show that the
separation of the high energy cosmic rays between protons and
nuclei can do this.

Besides the pion and $e^+e^-$ productions as in the proton case,
CMB photons can also disintegrate cosmic ray nuclei with energy of
the order $10^{19}$~eV and above through photonuclear processes
such as ${\mathrm{A}}+\gamma \to N+ {\mathrm{(A-1)}}$
\cite{GZK,Ste69}. There are similar processes for the interaction
between C$\nu$B neutrinos and cosmic ray nuclei. Taking helium
nuclei as an example, there are processes
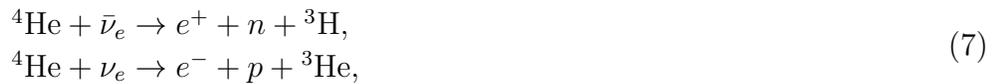
\begin{equation}
\begin{array}{ll}
^4{\mathrm{He}}+{\bar{\nu}_e}\to e^{+}+n+ {^3{\mathrm{H}}}, \\
^4{\mathrm{He}}+{{\nu}_e}\to e^{-}+p+ {^3{\mathrm{He}}},
\end{array}
\end{equation}
which can be expressed in a general form as
\begin{equation}
A+{\bar{\nu}_e}({{\nu}_e}) \to e^{\pm}+N+(A-1).
\end{equation}
Thus we get the threshold energy for the cosmic ray nuclei
\begin{equation}
E_{A}^{{\nu}_e}=\frac{S-m_A^2}{2\epsilon_{\nu}(1+\left|{\mathbf
k}\right|/\epsilon_{\nu})}=\frac{(m_{A-1}+m_N+m_e)^2-m_A^2}{2\epsilon_{\nu}(1+\left|{\mathbf
k}\right|/\epsilon_{\nu})}. \label{EAnu}
\end{equation}
If C$\nu$B neutrinos can cause an observable effect to the cosmic
ray protons by $p+{\bar{\nu}_e}\to e^+ +n$ at threshold energy
$E_{p}^{\nu_e}$, i.e., producing a ``knee" above the threshold and
a ``bump" below in the cosmic ray proton spectrum, then these
neutrinos should also produce a ``knee" in the spectrum of cosmic
ray A nuclei above $E_{A}^{{\nu}_e}$. There is no corresponding
``bump" below the threshold, as the cosmic ray A nuclei with
higher energies will be disintegrated into cosmic ray A-1 nuclei
or others. Comparing eqs.~(\ref{Epnu}) and (\ref{EAnu}), we get
the relation between the threshold energies
\begin{equation}
\frac{E^{{\nu}_e}_{p}}{E^{{\nu}_e}_{A}}=\frac{\left(m_n+
m_{e}\right)^2-m_p^2}{(m_{A-1}+m_N+m_e)^2-m_A^2}.
\end{equation}
Thus systematic measurements of the spectra of cosmic rays for
protons and nuclei can test whether process (\ref{pnu}) is an
observable effect in the cosmic rays or not. As there could be
various kinds of nuclei in the cosmic rays, such as
$^{12}\mathrm{C}$ and $^{56}\mathrm{Fe}$, the corresponding
thresholds can be exactly predicted with well established
knowledge of nuclear physics. It should be able to measure or
constrain the neutrino clustering factor $c_{\nu}$, provided that
the cosmic rays can be separated in composition between protons
and nuclei.

Once the threshold positions can be optimistically identified, we
can also use these positions for a direct measurement of the
neutrino mass, e.g., from eq.~(\ref{Epnu}) we get
\begin{equation}
m_{{\nu}_e}=\frac{\left(m_n+
m_{e}\right)^2-m_p^2}{2E^{{\nu}_e}_{p}}.
\end{equation}
Taking the starting point of the ``knee" in all particle cosmic
ray spectrum at energy $4\times 10^{15}$~eV as $E^{{\nu}_e}_{p}$,
we get $m_{{\nu}_e}=0.42$~eV, which is within available
experimental constraints \cite{triton} and is also consistent with
the estimate by Wigmans~\cite{Wig00}. One may naturally expect
that protons are dominated in the measured all-particle cosmic
rays around the knee, however, a claim of a so large effect of the
knee as arising from C$\nu$B neutrinos \cite{Wig00} seems
extraordinary. The origin of the knee in the all-particle spectrum
remains an open problem and there are also other novel
explanations \cite{KASCADE}. It would be safe to check the
corresponding knee in the energy spectrum for each species of
cosmic ray nuclei. Assuming $m_{{\nu}_e}=0.42$~eV, we get the
ratios of the threshold energies for p, $^4{\mathrm{He}}$,
$^{12}{\mathrm{C}}$, and $^{56}{\mathrm{Fe}}$ cosmic rays,
\begin{equation}
\begin{array}{ll}
& E^{{\nu}_e}_{p}: E^{{\nu}_e}_{^4{\mathrm{He}}}:
E^{{\nu}_e}_{^{12}{\mathrm{C}}}: E^{{\nu}_e}_{^{56}{\mathrm{Fe}}}
\\ \approx & 4 \times 10^{15}: 1.8 \times 10^{17}: 4.4 \times 10^{17}: 1.3 \times 10^{18}.
\end{array}
\label{Eratio}
\end{equation}
Adopting a different value of $m_{{\nu}_e}$ will change these
threshold positions, but do not change the ratios. Also the above
positions should be considered as a rough estimate, as detailed
consideration may change the position somewhat, for example,
contribution from processes such as
\begin{equation}
\begin{array}{ll}
^{56}{\mathrm{Fe}}+{\bar{\nu}_e}  \to e^{+}+ {^{56}{\mathrm{Mn}}},\\
^{56}{\mathrm{Fe}}+{{\nu}_e} \to e^{-}+ {^{56}{\mathrm{Co}}},
\end{array}
\end{equation}
where $^{56}{\mathrm{Mn}}$ and $^{56}{\mathrm{Co}}$ are unstable
nuclei, may produce lower threshold positions for
$^{56}{\mathrm{Fe}}$ at around $5 \times 10^{17}$~eV.

Balloon and satellite investigations can provide direct
measurements of the cosmic ray spectrum for each species of nuclei
respectively. Unfortunately, the available measurements can only
reach to the energy scale of $10^{15}$~eV \cite{Wef03}, still $1
\to 3$ orders below the energy scale for our purpose. The
available measurements of cosmic rays with energies ranging from
$10^{14}$ to a few $10^{20}$~eV are constructed from extensive air
showers \cite{EAS}. The energy and composition of the primary
cosmic rays are not directly measured, but can be reconstructed
from the air shower profiles such as electron and muon size
distributions \cite{KASCADE,Akeno}.

There have been experimental attempts to de-convolute the
all-particle energy spectrum around the knee into four individual
primary mass groups of p, He, C, and Fe by the KASCADE
collaboration \cite{KASCADE}. The mass information of primary
extensive air showers is deduced by the secondary electron and
muon shower size distributions. Each of these preliminary energy
distributions of the four mass groups exhibits a knee like
structure with an increase of their knee positions for heavier A
nuclei. It is interesting to note that the observed positions are
in agreement with the above predictions, as can be checked from
the KASCADE results \cite{KASCADE}. It is also possible that the
proton knee starts at $(1\to 2)\times 10^{15}$~eV \cite{Huang03},
which corresponds to the neutrino mass $m_{{\nu}_e}=0.84 \to
1.68$~eV, so that all corresponding knee positions from
Eq.~(\ref{Eratio}) are shifted by a factor $2\to 4$ to the lower
energy side. Considering that there are large uncertainties in
both theoretical predictions and experimental reconstructions, we
may take the KASCADE observation as possible support for a large
neutrino clustering factor to be measurable. Of course, more
precision and reliable measurements of detailed energy spectra of
cosmic ray protons and nuclei are still necessary, before one can
draw a conclusion to confirm or rule out such a situation.

The relative abundances of cosmic ray nuclei may show more
significant signature than the isolated spectrum for each species
of nuclei. Thus we may have chance to detect and constrain the
C$\nu$B with a smaller clustering factor. For example, the ratio
of $^3{\mathrm{He}}/{^4{\mathrm{He}}}$ is of the order $10^{-4}$
in the universe from primordial nucleosynthesis \cite{BBN}. If
this feature is also true at the source for very high energy
cosmic rays, then interaction of abundant cosmic ray
$^4{\mathrm{He}}$ nuclei with cosmic neutrinos, i.e.,
$^4{\mathrm{He}}+{{\nu}_e}\to e^{-}+p+ {^3{\mathrm{He}}}$, will
produce secondary cosmic ray $^3{\mathrm{He}}$ nuclei. The energy
of the produced $^3{\mathrm{He}}$ can be exactly calculated and it
should be below and around the $^4{\mathrm{He}}$ threshold energy.
This will cause a significant ``bump" in the ratio of
$^3{\mathrm{He}}/{^4{\mathrm{He}}}$ around the corresponding
energy position and a ``tail" below, in case the neutrino
clustering factor is of the order $10^{9}$ or larger.
Contributions from CMB photon disintegration of abundant cosmic
ray nuclei should happen at higher energy scale, thus do not
diminish the signature due to interaction of the cosmic ray nuclei
with cosmic neutrinos. One may also search for a pair of nuclei
with more significant difference in their relative abundances.
This can reduce the observable neutrino clustering factor to even
much smaller cases.

The abundance of cosmic ray nuclei can show more significant
effect in case with a higher neutrino clustering factor. For
example, the number of less abundant nuclei will be enhanced below
and around the knee as secondaries resulting from collisions of
the primary abundant cosmic ray nuclei with C$\nu$B neutrinos. We
can use this aspect as a further signature for neutrino
clustering, in addition to the knee onset position for each
species of cosmic ray nuclei. Other models for the knee may also
produce an increase of knee position for cosmic ray nuclei with
larger $A$ \cite{KASCADE}, and it is rather difficult to make a
distinction for the knee positions with charge number $Z$
dependence or nuclear number $A$ dependence \cite{Mau03}. But the
energy dependence in the relative abundances of cosmic ray nuclei
should be different in different models. Thus the cosmic ray
spectra of protons and nuclei with detailed energy dependence and
abundance information can provide decisive detection or constraint
on the cosmic neutrinos. Indeed, the isotopic ratio of
$^3{\mathrm{He}}/{^4{\mathrm{He}}}$ has been measured in balloon
experiments \cite{He3}, and this ratio is found significantly
enhanced compared to that from nucleosynthesis, with also a trend
to increase as energy increases. However, the energies of those
cosmic rays only range around $10^8$ to $10^{10}$~eV, rather low
to conclude as the ``tail" resulting from the mechanism we
suggested.

In comparison with Wigmans' proposal to attribute the knee in high
energy cosmic ray spectrum as arising from C$\nu$B neutrinos, we
provide a practical relation, i.e., Eq.~(\ref{Eratio}), between
the ``knees" caused by interactions of C$\nu$B neutrinos with the
protons and specific species of nuclei\footnote{The discussion on
the reaction of C$\nu$B neutrinos with $\alpha$ particles by
Wigmans is not appropriate from nuclear physics, as the
corresponding reaction should be Eq.(7) in this paper}. Such
relations can provide a crucial test of the idea to attribute the
existing knee as a signal for the interaction between the protons
with C$\nu$B neutrinos. We also propose a practical method to
measure the isotopic ratio such as
$^3{\mathrm{He}}/{^4{\mathrm{He}}}$, so that we may have chance to
detect and constrain the C$\nu$B with a smaller clustering factor.

In summary, we proposed a method to investigate the scenario that
cosmic relic neutrinos are highly clustered around stars and
galaxies, or dark-matter clusters, rather than uniformly
distributed in the universe. High energy cosmic ray protons and
nuclei should interact with these cosmic neutrinos through their
propagation in the universe. There should be observable signature
in cosmic ray energy spectra of protons and nuclei, for a neutrino
clustering factor beyond $10^{13}$.  The onset positions of the
knee signature should be closely related between protons and
nuclei, and there seems some support from existing experiments. We
have also indicated that a smaller neutrino clustering factor can
be detected or constrained by an increase of secondary less
abundant cosmic ray nuclei produced from interaction of the
corresponding primary abundant cosmic ray nuclei with the cosmic
neutrino background. The neutrino mass can also be directly
measured from the knee onset positions reflecting interactions of
the high energy cosmic ray protons and nuclei with cosmic
neutrinos. Thus systematic investigations of high energy cosmic
ray protons and nuclei are practical to reveal or constrain the
cosmic neutrino clustering and neutrino properties. Precision
measurements are required for this purpose in balloon, satellite,
and extensive air shower experiments of high energy cosmic rays
with explicit nuclei composition information included.

%\begin{acknowledgments}

{\bf Acknowledgments: } This work is partially supported  by the
Taiwan CosPA Project funded by the Ministry of Education
(89-N-FA01-1-0 up to 89-N-FA01-1-5). It is also supported by the
National Natural Science Foundation of China under Grant
Nos.~10025523 and 90103007.

\end{document}